\documentclass[doublecol,linenumbers]{epl2} 
\usepackage{graphicx}
\usepackage{textcomp}
\usepackage{amsmath,amssymb}
\usepackage{color}
\usepackage{epsfig}
\usepackage{subfigure}
\usepackage{epstopdf}
\usepackage{amsthm}
\usepackage{bm}
\usepackage{float}
\begin{document}
\title{Schwinger's Model of Angular Momentum with GUP}
\author{ Harshit~Verma\inst{1}, Toshali~Mitra\inst{2} \and Bhabani~Prasad~Mandal\inst{2}}
\shortauthor{H. Verma \etal}
\institute{                    
 \inst{1} Department of Physics, Indian Institute of Technology (Banaras Hindu University), Varanasi, India\\
 \inst{2} Department of Physics, Institute of Science, BHU, Varanasi, India }
\abstract{ We study the generalized uncertainty principle (GUP) modified simple harmonic oscillator (SHO) in the operator formalism by considering the appropriate form of the creation and annihilation operators $A, A^\dagger$. The angular momentum algebra is then constructed using Schwinger's model of angular momentum with two independent GUP modified SHOs. With the GUP modified angular momentum algebra, we discuss  coupling of angular momentum for a two-particle composite system. Further, we calculate the Clebsch-Gordan (CG) coefficients for a two-particle system explicitly. Our results show that the CG coefficients do not receive any corrections upto quadratic GUP.}

\maketitle
\section{Introduction}

Using the Heisenberg uncertainty principle, we may infer that for detecting an arbitrarily small scale, tools of sufficiently high energy (high momentum) are required. Thus, essential quantum phenomena must be probed at high energies. Interestingly, gravity also becomes important at high energies. Therefore, we must resort to quantum gravity theories when trying to address fundamental areas of physics. One of the major aims of physics has always been to reconcile gravity and quantum mechanics in the same framework and possibly arrive at a consistent and complete theory of quantum gravity. In this regard, many theories such as string theory, canonical quantum gravity etc. have been formulated that have emerged as strong candidates for the same.\\ 
Interestingly, all theories of quantum gravity and black hole physics predict the existence of minimum length scale called the Planck length ($l_{pl} \approx 10^{-35}m$) \cite{s8,s9,s10,s11,s12,s4,PP3}. This leads to the modification of usual Heisenberg uncertainty principle. The new uncertainty relation that we get from these theories is called the generalized uncertainty principle (GUP)\cite{s11, Amelino,Garay,Maggi,Scar,brs}. The GUP as derived from string theory and black hole physics is given as :
\begin{eqnarray} \label{Eq.1}
\Delta p \Delta x \geq \frac{\hbar}{2} \bigg[ 1+\beta_0 \frac{l_{pl}^2}{\hbar^2} \Delta p ^2 \bigg]
\end{eqnarray}
 where, $l_{pl}=\sqrt{\frac{G\hbar}{c^3}}=10^{-35}$m is the Planck length and $\beta_0$ is a constant, assumed to be of the order of unity. Evidently, the new second term on the RHS of the above equation is important only when $\Delta x \approx l_{pl}$ or $\Delta p \approx p_{pl} \approx 10^{16}$ TeV/c (Planck momentum) i.e. at very high energies / small length scales. \\
The commutator algebra between $x_i, p_j$ from string theory and double special relativity can be reconciled into \cite{Scar,kemph}:
\begin{eqnarray}\label{Eq.2}
[x_i, p_j] = i\hbar\bigg[ \delta_{ij}- \alpha \bigg(p\delta_{ij}+\frac{p_i p_j}{p}\bigg)+\alpha^2 (p^2 \delta_{ij}+3 p_i p_j)\bigg]
\end{eqnarray}
where $\alpha=\frac{\alpha_0}{M_{pl}c}=\frac{\alpha_0 l_{pl}}{\hbar}$. Here $\alpha_0$ is normally assumed to be of the order of unity. Both $\Delta x_{min}$ and $\Delta p_{min}$ arise as an implication of the above algebra. In one dimension, it leads to the following form of GUP \cite{PP3}:
\begin{eqnarray}\label{Eq.3}
\Delta x \Delta p \geq \frac{\hbar}{2}[1-2\alpha \langle p \rangle + 4 \alpha^2 \langle p^2 \rangle ]~.
\end{eqnarray}
The above mentioned modified Heisenberg algebra can be derived if we represent the position and momentum operators in the following way \cite{PP3}: 
\begin{eqnarray}\label{Eq.4}
x_i  = x_{0i} ,~~p_i = p_{0i}(1-\alpha p_0 + 2\alpha^2 p_0^2)
\end{eqnarray}
where $x_{0i}$ and $p_{0j}$ satisfy the usual canonical commutation relations $[x_{0i},p_{0j}] = i\hbar \delta_{ij}$ and are interpreted as low energy position and momenta respectively. On the other hand $x_i$ and $p_i$ are the position and  momentum at high energy respectively. Since $\alpha_0$ is of the order of unity, $\alpha$ dependent terms become important only at energy (momentum) scales comparable to Planck energy (momentum) scale.
The GUP and its effects have been studied extensively over the past decade \cite{s15}-\cite{s5}. Apart from its important implications in cosmology \cite{cos1,cos2,cos3} and black hole physics \cite{bh1,bh2}, Klein-Gordan \cite{chen} and Dirac equations \cite{oth,dirac} have also been modified to include quantum gravity effects. Various non-relativistic quantum systems such as particle in a box, SHO, Landau levels \cite{PP3,s6}, coherent and squeezed states \cite{cs,cs1,cs2,cs3,s3}, PT symmetric non-Hermitian systems \cite{oth2,oth3,oth4} have been investigated in the framework of GUP to note the effects of quantum gravity in those. A particularly interesting case with a probable observable effect has been developed in \cite{PP3} by studying variations in current of scanning tunnel microscope (STM) when tunnelling phenomenon is subject to GUP corrections.\\
Recently, GUP modified angular momentum algebra has been constructed  and the corrections in CG coefficients due to GUP are obtained \cite{PP2}.
However, their results are inconsistent (as mentioned by the authors themselves) as the CG coefficients obtained by applying $J_-$ to the highest $J_m$ value state and is different from that of obtained by applying $J_+$ to the lowest $J_m$ value state. This motivate us to revisit the coupling  angular momentum in the framework of GUP.
In this work, we construct GUP modified angular momentum algebra by using Schwinger's model of angular momentum (SMAM)\cite{s1}. In the first step we develop the operator formulation for the GUP modified SHO. By modifying the creation/annihilation operators appropriately, we write the GUP modified SHO hamiltonian as $\frac{\hbar\omega}{2} [A^\dagger A + A A^\dagger]$ with the commutation relation between $A$ and $A^\dagger$ as $[A, A^\dagger] = 1+\mathcal{O}(\alpha)$. Angular momentum algebra is then constructed by considering two such independent oscillators. This GUP modified algebra is then used to study two-particle angular momentum system. We explicitly calculate the CG coefficients. We found CG coefficients obtained by applying $J_-$ and $J_+$ are same unlike the  case in Ref. \cite{PP2}. We further found  that CG coefficients remain unaffected in upto quadratic GUP, even though the angular momentum algebra gets modified. This is very interesting result in the GUP framework. However, it is not surprising as the GUP modified angular momentum algebra can be mapped to usual angular momentum algebra by redefining the Planck constant.\\
We now present the plan of the paper. In the next section, we develop the complete GUP modified formalism of SHO in the operator formulation and then use SMAM to derive GUP modified angular momentum algebra in Sec. 3. We discuss angular momentum coupling in a composite two-particle system in the GUP framework in Sec.4. In Sec. 5, CG coefficients are calculated explicitly for two particles system. Finally, Sec. 6 is for the concluding remarks. 
%\section{Results and Discussion}
\section{GUP modified Harmonic Oscillator}
In this section, we discuss the operator formulation of SHO in the framework of GUP  using both high energy and low energy momentum variables. 
\subsection{Using low energy Momentum}
The corrections to the harmonic oscillator model due to GUP have been addressed mostly using perturbation theory\cite{PP3} and also treated  in phase space \cite{s5}. In the present work, we formulate the operator formalism for the GUP corrected SHO. The GUP modified harmonic oscillator hamiltonian is given as:
\begin{eqnarray}\label{Eq.5}
H = \hbar \omega\bigg(\frac{x^2}{x^{\prime 2}}+ \frac{p^2}{p^{\prime 2}}\bigg) 
\end{eqnarray} 
where x and p are the high energy position and momentum operators.
This GUP corrected Hamiltonian can be written in terms of low energy operators upto $\mathcal{O}(\alpha^2)$ as :
\begin{eqnarray}\label{Eq.6}
H = H_0 + H_1 = \hbar\omega\bigg(\frac{x_0^2}{x^{\prime 2 }} +\frac{p_0^2}{p^{\prime 2}}\bigg)+\hbar\omega\bigg(-\frac{2\alpha p_0^3}{p^{\prime 2}}+\frac{5\alpha^2p_0^4}{p^{\prime 2}}\bigg)
\end{eqnarray}
where,
\begin{eqnarray}\label{Eq.7}
H_0 = \hbar\omega\bigg(\frac{x_0^2}{x^{\prime 2}}+\frac{p_0^2}{p^{\prime 2}}\bigg) ,\qquad H_1 =  \hbar\omega\bigg(-\frac{2\alpha p_0^3}{p^{\prime 2}}+\frac{5\alpha^2p_0^4}{p^{\prime 2}}\bigg).
\end{eqnarray}
Here, $H_1$ is the correction due to GUP, $x^\prime=\sqrt{\frac{2\hbar}{m\omega}}$ and $p^\prime=\sqrt{2m\hbar\omega}$ and $x_0$ and $p_0$ are usual low energy position and momentum operators respectively.\\
Now  we modify the usual creation ($a^\dagger$) and annihilation ($a$) operators of 1-d harmonic oscillator in the following manner:
\begin{eqnarray}\label{Eq.8}
A^\dagger &=&\frac{x_0}{x^\prime}- i \frac{p_0}{p^\prime}(1-\alpha p_0 + 2 \alpha^2 p_0^2) \nonumber\\ &+& \alpha\beta_1^\dagger(x_0,p_0) + \alpha^2 \beta_2^\dagger(x_0,p_0)\\
\label{Eq.9}
A &=&\frac{x_0}{x^\prime}+ i \frac{p_0}{p^\prime}(1-\alpha p_0 + 2 \alpha^2 p_0^2)\nonumber\\ &+& \alpha\beta_1(x_0,p_0) + \alpha^2 \beta_2(x_0,p_0)
\end{eqnarray}
in which, the usual creation/annihilation operators $a^\dagger=\frac{x_0}{x^\prime}-i\frac{p_0}{p^\prime}$ and $a=\frac{x_0}{x^\prime}+i\frac{p_0}{p^\prime}$ are modified by replacing $p = p_0(1-\alpha p_0 +2\alpha^2 p_0^2)$ and $x=x_0$ and including corrections to the order $\alpha$ ($\beta_1(x_0,p_0)$) and $\alpha^2$ ($\beta_2(x_0,p_0)$).\\
We demand the Hamiltonian in Eq.\ref{Eq.6} to be in the following form:
\begin{eqnarray}\label{Eq.10}
H = \frac{\hbar\omega}{2} (A A^\dagger + A^\dagger A)
\end{eqnarray}
by constructing appropriate $\beta_1$ and $\beta_2$ as given by:
\begin{eqnarray}\label{Eq.11}
\beta_1 &= i a p^\prime ,\qquad &\beta_1^\dagger = -i a^\dagger p^\prime \\
\label{Eq.12}
\beta_2 &= p_0^2 - \frac{p^{\prime 2}}{2}a ,\qquad &\beta_2^\dagger = p_0^2 - \frac{p^{\prime 2}}{2} a^\dagger ~.
\end{eqnarray}
We now calculate the commutator $[A,A^\dagger]$, which comes out upto $\mathcal{O}(\alpha^3)$ as:
\begin{eqnarray}\label{Eq.13}
[A,A^\dagger] &=& 1- 2\alpha p_0 + 6\alpha^2 p_0^2 \nonumber \\
 & \equiv & 1-\mathcal{C}.
\end{eqnarray}
where $\mathcal{C}$ is the correction due to GUP.
In absence of GUP corrections, i.e. when $\alpha=0$, this formulation reduces to the usual operator formulation of 1-d SHO.
We now wish to redefine the creation/annihilation operators in GUP corrected formalism. For this purpose, we henceforth consider the GUP corrections as $\langle \mathcal{C} \rangle $ rather than $\mathcal{C} $. This has been considered under the assumption that, in experimental conditions, GUP correction to any operator manifests as expectation value. In the next section, we will use the GUP modified creation/annihilation operators to generate the angular momentum algebra. Therefore, $\langle \mathcal{C} \rangle$ is the averaged value of GUP corrections and our assumption stands correct (Ref. [Eq.36] in \cite{PP2}). Hence, Eq. \ref{Eq.13} is rewritten as:
\begin{eqnarray} \label{Eq.14}
[A, A^\dagger] = 1- \langle \mathcal{C} \rangle
\end{eqnarray}
We may now assume two new operators $\tilde{A}$ and $\tilde{A^\dagger}$ such that,
\begin{eqnarray} \label{Eq.15}
[\tilde{A},\tilde{A^\dagger}]=1
\end{eqnarray}
Here, $\tilde{A}=A/\sqrt{1-\langle \mathcal{C} \rangle}$ and $\tilde{A^\dagger}=A^\dagger/\sqrt{1- \langle \mathcal{C} \rangle}$.
If $\vert 1 \rangle, \vert 2 \rangle ,..., \vert N\rangle $ are eigenvectors of Hamiltonian in Eq.\ref{Eq.10}, then, in a general momentum regime, we can write,
\begin{eqnarray}
\tilde{A}|N\rangle &=& \sqrt{N}~|N-1\rangle \\ 
\tilde{A^\dagger}|N\rangle &=& \sqrt{N+1}~|N+1\rangle
\end{eqnarray}
or equivalently,
\begin{eqnarray}\label{Eq.19}
A|N\rangle &=& \sqrt{1- \langle \mathcal{C}\rangle}~\sqrt{N}~|N-1\rangle \\
\label{Eq.20}
A^\dagger|N\rangle &=& \sqrt{1- \langle\mathcal{C}\rangle}~\sqrt{N+1}~|N+1\rangle
\end{eqnarray}
where, $\langle \mathcal{C} \rangle = 2\alpha \langle p_0 \rangle - 6 \alpha^2 \langle p_0^2 \rangle $ such that commutation relation in Eq. \ref{Eq.14} is satisfied. In the above equations and in the later sections, we have expressed GUP effects solely in terms of expectation value of $\mathcal{C} $ based on aforementioned arguments.
\subsection{Using high energy momentum}
Now, we would like to show that conclusion of the previous section can also be achieved while remaining in the high momentum regime (i.e. GUP modified momentum). The GUP modified creation/annihilation operators $A$ and $A^\dagger$ are defined in terms of high energy variables (x,p) as:
\begin{eqnarray}\label{Eq.21}
A^\dagger &=&\frac{x}{x^\prime}- i \frac{p}{p^\prime}\\
\label{Eq.22}
A &=&\frac{x}{x^\prime}+ i \frac{p}{p^\prime}
\end{eqnarray}
where x and p are in high momentum regime and $x^\prime=\sqrt{\frac{2\hbar}{m\omega}}$ and $p^\prime=\sqrt{2m\hbar\omega}$.
Then the GUP modified Hamiltonian is written as $H = \frac{\hbar\omega }{2}[A A^\dagger + A^\dagger A]$ and the commutator $[A, A^\dagger]$ in the regime is obtained as:
\begin{eqnarray}\label{Eq.23}
[A, A^\dagger] = 1 - 2 \alpha p + 4 \alpha^2 p^2
\end{eqnarray}
which is exactly same as in Eq.\ref{Eq.13} when p is expressed in terms of low energy variables as $p = p_0 (1- \alpha p_0 + 2\alpha^2 p_0^2)$.
In general momentum regime, similar to the discussion for low energy momentum, we have,
\begin{eqnarray}\label{Eq.26}
A|N\rangle &=& \sqrt{1- \langle\mathcal{C}\rangle}~\sqrt{N}~|N-1\rangle \\
\label{Eq.27}
A^\dagger|N\rangle &=& \sqrt{1- \langle\mathcal{C}\rangle}~\sqrt{N+1}~|N+1\rangle
\end{eqnarray}
where, $\langle \mathcal{C} \rangle = 2\alpha \langle p\rangle - 4 \alpha^2 \langle p^2\rangle$.

\section{GUP modified angular momentum algebra using Schwinger's Model of Angular Momentum}
In this section, we would like to calculate the GUP modified angular momentum algebra by considering two such GUP modified harmonic oscillators. Two uncoupled SHOs (we call them type 1 and 2) are considered with number states denoted as $n_1$ and $n_2$ with their usual independent algebra. Then the general eigenket is constructed in SMAM as:
 \begin{eqnarray}\label{Eq.28}
 N_{1,2} \vert n_1 ,n_2 \rangle &=& n_{1,2} \vert n_1 ,n_2 \rangle \\
\label{Eq.29} 
 a_1^\dagger \vert n_1 ,n_2 \rangle &=& \sqrt{n_1 + 1} \vert n_1 +1 ,n_2 \rangle\\
 \label{Eq.30}
 a_1 \vert n_1 ,n_2 \rangle &=& \sqrt{n_1} \vert n_1 -1 ,n_2 \rangle
  \end{eqnarray}
and similar relations are obtained when  $a_2^\dagger $ and $a_2$ act on these states.
  Next, in SMAM the following angular momentum operators are constructed \cite{s1}:
\begin{eqnarray}\label{Eq.31}
L_+\vert n_1,n_2\rangle & \equiv & \hbar~a_1^\dagger a_2\vert n_1,n_2\rangle \nonumber\\ &=& \hbar[n_2(n_1+1)]^{1/2}\;\vert n_1+1,n_2-1\rangle\\
\label{Eq.32}
L_-\vert n_1,n_2\rangle & \equiv & \hbar~a_2^\dagger a_1\vert n_1,n_2\rangle \nonumber\\ &=& \hbar[n_1(n_2+1)]^{1/2}\;\vert n_1-1,n_2+1\rangle\\
\label{Eq.33}
L_z\vert n_1,n_2\rangle & \equiv & \frac{\hbar}{2}(N_1-N_2)\vert n_1,n_2\rangle \nonumber\\ &=& \frac{\hbar}{2}(n_1-n_2)\vert n_1,n_2\rangle
\end{eqnarray}
which satisfy the angular momentum algebra $[L_z, L_\pm ] = \pm \hbar L_\pm$, $[L_+, L_-] = 2\hbar L_z$.
We now make a canonical transformation to $n_1$ and $n_2$ in the above.
\begin{eqnarray}\label{Eq.34}
n_1\to l+m,\qquad n_2\to l-m ~.
\end{eqnarray}
Then, the Eq.\ref{Eq.31}, \ref{Eq.32}, \ref{Eq.33} reduce to the usual form of angular momentum algebra as shown in Eq.\ref{Eq.35}, \ref{Eq.36}, \ref{Eq.37}
\begin{eqnarray}\label{Eq.35}
L_+\vert l+m,l-m\rangle &=& \lambda_{+}\vert l+m+1,l-m-1\rangle\\
\label{Eq.36}
L_-\vert l+m,l-m\rangle &=&\lambda_{-}\vert l+m-1,l-m+1\rangle\\
\label{Eq.37}
L_z\vert l+m,l-m\rangle &=& \hbar~m\vert l+m,l-m\rangle,
\end{eqnarray}
where $\lambda_\pm =  \hbar~[(l \mp m)(l\pm m+1)]^{1/2}$ and the eigenvalues of the quadratic operator $\vec{L}^2$ become 
\begin{eqnarray}\label{Eq.38}
\vec L^2\vert l+m,l-m\rangle&=& \hbar^2 l(l+1)\vert l+m,l-m\rangle ~.
\end{eqnarray}
Thus, we see that angular momentum algebra is constructed using two SHO algebra when eigenstates are re-labelled by replacing $|l+m,l-m\rangle$ by $|l,m\rangle$.
We follow the same technique now for GUP modified harmonic oscillator in the high energy momentum regime. Repeating the same steps of SMAM (Eq.\ref{Eq.35},\ref{Eq.36},\ref{Eq.37},\ref{Eq.38}) with GUP modified SHO algebra (Eq. \ref{Eq.23}, \ref{Eq.26},\ref{Eq.27}), we obtain,
%\begin{widetext}
\begin{eqnarray}\label{Eq.39}
L_+\vert l,m\rangle &=& \hbar~(1- \langle \mathcal{C} \rangle)~\lambda_+\vert l,m+1\rangle\\
\label{Eq.40}
L_-\vert l,m\rangle &=& \hbar~(1- \langle \mathcal{C} \rangle)~\lambda_-\vert l,m-1\rangle\\
\label{Eq.41}
L_z\vert l,m\rangle &=& \hbar~(1- \langle \mathcal{C} \rangle)~m~\vert l,m\rangle\\
\label{Eq.42}
\vec L^2\vert l,m\rangle &=& \hbar^2~(1- \langle \mathcal{C} \rangle)^2~l(l+1)~\vert l,m\rangle ~.
\end{eqnarray}
%\end{widetext}
The form $L_x$, $L_y$ are then constructed using: 
\begin{eqnarray}\label{Eq.43}
L_x = \frac{L_+ + L_-}{2} ,\qquad L_y = \frac{L_+ - L_-}{2 i} ~.
\end{eqnarray}
We now find the various commutation relations between $L_\pm$, $L_z$, $L^2$:
%\begin{widetext}
\begin{eqnarray}\label{Eq.44}
\left[L_z , L_+\right]\vert l,m\rangle &=& (L_z L_+ - L_+ L_z)\vert l,m \rangle\nonumber\\  
%&=& \hbar^2 (m+1)(1-\langle \mathcal{C} \rangle)^2[(l-m)(l+m+1)]^{1/2} \vert l, m+1\rangle \nonumber \\
%&-& \hbar^2 (m)(1-\langle \mathcal{C} \rangle)^2[(l-m)(l+m+1)]^{1/2} \vert l, m+1\rangle \nonumber \\
&=& \hbar^2 (1-\langle \mathcal{C} \rangle)^2\lambda_+ \vert l, m+1\rangle ~.
\end{eqnarray}
%\end{widetext}
Thus, we have,
\begin{eqnarray}\label{Eq.45}
[L_z, L_+ ]\vert l,m\rangle = \hbar (1-\langle \mathcal{C} \rangle) L_+ \vert l,m\rangle
\end{eqnarray}
Similarly, on calculation using Eq.\ref{Eq.39},\ref{Eq.40},\ref{Eq.41},\ref{Eq.42}, we get the following :
\begin{eqnarray}\label{Eq.46}
\left[L_z , L_-\right]\vert l,m\rangle &=& -\hbar (1-\langle \mathcal{C} \rangle) L_- \vert l,m\rangle\\
\label{Eq.47}
\left[L^2 , L_\pm\right]\vert l,m\rangle &=& 0\\
\label{Eq.48}
\left[L_+ , L_-\right]\vert l,m\rangle &=& 2\hbar L_z (1-\langle \mathcal{C} \rangle)\vert l,m\rangle\\
\label{Eq.49}
\left[L_x, L_y \right]\vert l,m\rangle &=& i \hbar L_z (1-\langle \mathcal{C} \rangle)\vert l,m\rangle .
\end{eqnarray}
Commutation relation given by Eq.\ref{Eq.47} implies that the ladder operator doesn't change the magnitude of total angular momentum ($l$). Also, due to occurrence of the correction factor term, the spacing between consecutive eigenvalues of $L_z$ changes.
%\begin{widetext}[!h]
\begin{eqnarray}\label{Eq.50}
L_z L_\pm |l, m\rangle = L_\pm [ L_z \pm \hbar(1- \langle \mathcal{C} \rangle )]|l, m\rangle  \nonumber \\
= \hbar (1- \langle \mathcal{C} \rangle )[m \pm 1] L_\pm|l,m\rangle
\end{eqnarray}
%\end{widetext}
Also, Eq. \ref{Eq.50} implies that $L_\pm |l,m\rangle $ is an eigenstate of $L_z$. Now, using,
\begin{eqnarray}
\label{Eq.51}
L_- L_+ &=& L^2 - L_z[L_z + \hbar(1 - \langle \mathcal{C} \rangle )] \nonumber\\ 
L_+ L_- &=& L^2 - L_z[L_z - \hbar(1 - \langle \mathcal{C} \rangle )]~,
\end{eqnarray}
we can calculate the norm of raised lowered states $L_\pm \vert l,m\rangle$ as,
\begin{eqnarray}\label{Eq.52}
||L_\pm | l, m\rangle ||^2 &=& \langle l, m | L_\mp L_\pm | l, m \rangle \nonumber\\
&=& \hbar^2 (1- \langle \mathcal{C} \rangle )^2 [l(l+1) - m(m \pm 1)] \nonumber\\
&\geq & 0~. 
\end{eqnarray}
Therefore, as in the usual quantum mechanics \cite{s1} we have limit on the values of $m$, which we denote $m_{max}$ (maximum m value or upper bound) and $m_{min}$ (minimum m value or lower bound).
Application of raising operator $L_+$ on state characterised by $m_{max}$ and lowering operator $L_-$ on state characterised by $m_{min}$ would both give zero i.e.
\begin{eqnarray}\label{Eq.53}
	L_+ |l, m_{max}\rangle = 0 \qquad L_- |l, m_{min}\rangle = 0 ~.
\end{eqnarray}
Clearly, we observe that if we consider $\hbar \rightarrow \hbar (1-\langle \mathcal{C} \rangle)$ in the usual QM, we get the GUP modified formalism, where $ \langle\mathcal{C}\rangle = 2\alpha \langle p \rangle - 4 \alpha^2 \langle p^2 \rangle$.

\section{Effect on two-particle system}
Till now all the algebra derived above involves single particle. In this section, the effect of GUP on multi-particle algebra will be examined. We will start with two-particle angular momentum algebra and the results can be extended for N-particle systems accordingly.

%\subsubsection{Addition of angular momentum}
Consider the addition of angular momentum including GUP for a system of two particles with $l_i$ and $m_i$ ($i=1,2$) ,the azimuthal and magnetic quantum numbers of the particles. The total angular momentum and the $ z-$component for the composite system is
\begin{eqnarray}\label{Eq.54}
J = j_1 + j_2 \qquad J_z = j_{1,z} + j_{2,z} ~.
\end{eqnarray} 
We have two  alternative choices of the basis for representing the combined system as
\begin{enumerate}
\item With simultaneous eigenkets of $J_1^2$, $J_2^2$, $J_{1,z}$, and $J_{2,z}$. denoted by $\vert j_1,j_2; m_1, m_2 \rangle $ or,
\item With simultaneous eigenkets of $J^2$, $J_1^2$, $J_2^2$, and $J_z$. denoted by $\vert j_1, j_2; j, m \rangle $
\end{enumerate}
We chose the later one and the combined state in Hilbert space is represented as,
\begin{eqnarray}\label{Eq.55}
| j_1, j_2; j,m \rangle &=& \sum_{m_1}\sum_{m_2} \vert j_1, j_2; m_1,m_2\rangle \langle j_1, j_2; m_1,m_2 |j_1 , j_2; j, m \rangle \nonumber \\
&\equiv & \sum_{m_1 + m_2 = m} C_{m_1, m_2, m}^{j_1, j_2, j} \vert j_1, j_2; m_1,m_2\rangle
\end{eqnarray}
where, $|j_1, j_2; m_1, m_2\rangle = |j_1 , m_1 \rangle \otimes |j_2 , m_2 \rangle $  and $C_{m_1, m_2, m}^{j_1, j_2, j} \equiv \langle j_1, j_2; m_1,m_2 |j_1 , j_2; j, m \rangle $ are the Clebsch-Gordan coefficients. The CG coefficients may be calculated by applying $J_- $. To discuss the addition of angular momentum for this composite system further, we need the following equations. Operating $ J_z $  on the combined state we obtain Eq.\ref{Eq.56}, \ref{Eq.57}.
\begin{widetext}[!h]
\vspace{-0.5cm}
\begin{eqnarray}\label{Eq.56}
J_z | j_1, j_2; j,m \rangle &=& \sum_{m_1}\sum_{m_2}[\langle j_1, j_2; m_1,m_2 |j_1 , j_2; j, m \rangle ] \hbar [(1-\langle C_1 \rangle ) m_1 |j_1 , j_2; m_1, m_2 \rangle +  (1-\langle C_2 \rangle ) m_2 |j_1 , j_2 ;m_1, m_2\rangle]\\
\label{Eq.57}
J_z | j_1, j_2; j,m \rangle & = & \sum_{m_1 + m_2 = m} C_{m_1, m_2, m}^{j_1, j_2, j} \hbar [(1-\langle C_1 \rangle ) m_1 + (1-\langle C_2 \rangle ) m_2]|j_1 , j_2 ;m_1, m_2\rangle ] ~.
\end{eqnarray} 
\vspace{-0.5cm}
\end{widetext}
The following point may be noted in analogy to ordinary quantum mechanics and as done in \cite{PP2},
\begin{eqnarray}\label{Eq.58}
|j_1 - j_2| \leq J \leq |j_1 + j_2| ~.
\end{eqnarray} 
Now, let us consider the ladder operator for two - particle system. 
\begin{eqnarray}
\label{Eq.59}
J_{\pm} = j_{1,\pm} + j_{2,\pm}
\end{eqnarray}
From the commutation relations Eq.\ref{Eq.45},\ref{Eq.46} we find that,
\begin{eqnarray}\label{Eq.60}
[j_{1,z} , j_{1,\pm}] = \pm \hbar j_{1,\pm} (1-\langle C_1 \rangle)\\
\label{Eq.61}
[j_{2,z} , j_{2,\pm}] = \pm \hbar j_{2,\pm} (1-\langle C_2 \rangle)\\
\label{Eq.62}
[j_{1,z} , j_{2,\pm}] = 0, \qquad [j_{2,z} , j_{1,\pm}] = 0
\end{eqnarray}
 From this one gets,
\begin{eqnarray}\label{Eq.63}
J_{z}J_{\pm}|j_1,j_2 ; m_1,m_2\rangle =& (J_{\pm}J_z \pm \hbar (j_{1,\pm}(1-\langle C_1 \rangle ) \nonumber\\
\pm & j_{2,\pm}(1-\langle C_2 \rangle ))) |j_1 , j_2 ; m_1 , m_2\rangle \nonumber
\end{eqnarray}
\begin{eqnarray}
= \hbar\big[ j_{1,\pm}[(1-\langle C_1 \rangle)(m_1 \pm 1) + (1-\langle C_2 \rangle) m_2] \nonumber \\
+ j_{2,\pm}[(1-\langle C_2 \rangle)(m_2 \pm 1) + (1-\langle C_1 \rangle) m_1] |j_1,j_2 ; m_1, m_2\rangle
\end{eqnarray}
where, $\langle C_1 \rangle, \langle C_2 \rangle $ are the correction factors due GUP  for $|j_1,j_2;m_1,m_2\rangle$ state for first and second particle respectively. Also, for any state expressed as in Eq.\ref{Eq.55}, we have found the operation of $J_z J_\pm$ on the RHS states. As for the LHS of Eq.\ref{Eq.55}, the solution is easy to calculate. Eq.\ref{Eq.63} shows that the RHS is no longer an eigenstate of ${J_z}$ , unlike the $\alpha = 0$ case.\\ 
The above discussions on two- particle system can be also generalised for a system of N-particles on similar lines as done above.

\section{Clebsch-Gordan Coefficients : Two Particles}

In this section we calculate the Clebsch-Gordan coefficients explicitly for two particles. We  start with the highest $m$ value state in the highest multiplet and apply $J_-$. which lowers the $m$ value. 
We now resort to the notation $|j_1, j_2; j, m \rangle \equiv |J,M \rangle $; $|j_1, j_2; m_1, m_2 \rangle \equiv |m_1; m_2 \rangle$ where $J=j, M=m $.
\begin{enumerate}
 \item For ${ \bf  J = J_{m} , M= J_{m} } $ \, ,  where $ J_m $ stands for $ J_{max} =j_1 + j_2$
\begin{eqnarray}\label{Eq.72}
|J_m , J_m \rangle = |m_1=j_1;m_2=j_2 \rangle ~.
\end{eqnarray}
The coefficient is $ \langle j_1; j_2~|~J_m, J_m \rangle = 1 $ by normalization.
\item For ${ \bf  J = J_{m} , M= J_{m} - 1 } $ :\\
By applying $J_-$ we  calculate,
\begin{eqnarray}\label{Eq.73}
|J_m, J_m - 1\rangle =& \frac{(1-\langle C_1 \rangle)}{(1-\langle C\rangle)} \frac{\sqrt{j_1}}{\sqrt{j_1+j_2}} |j_1-1;j_2 \rangle \nonumber \\
+& \frac{(1-\langle C_2 \rangle)}{(1-\langle C\rangle)} \frac{\sqrt{j_2}}{\sqrt{j_1+j_2}} |j_1;j_2-1 \rangle  ~.
\end{eqnarray}
Normalization of $|J_m, J_m - 1\rangle $ in Eq.\ref{Eq.73} provides,
\begin{eqnarray}\label{Eq.74}
\bigg[ \frac{(1-\langle C_1 \rangle)^2}{(1-\langle C\rangle )^2}\bigg] \bigg[\frac{j_1}{j_1+j_2}\bigg] + \bigg[ \frac{(1-\langle C_2 \rangle )^2}{(1-\langle C\rangle )^2} \bigg] \bigg[\frac{j_2}{j_1+j_2} \bigg] \nonumber \\
= 1 ~.
\end{eqnarray}
\item For ${ \bf  J = J_{m} , M= J_{m}-2 } $ :\\
Applying $J_-$ again we obtain Eq. \ref{Eq.75}
\begin{widetext}[!h]
\vspace{-0.5cm}
\begin{eqnarray}\label{Eq.75}
|J_m, J_m-2 \rangle &=& \bigg[ \frac{(1-\langle C_1 \rangle)^2}{(1-\langle C\rangle )^2}\bigg] \sqrt{\frac{j_1(2 j_1 -1)}{(j_1 + j_2)(2(j_1 + j_2)-1)}} | j_1-2; j_2\rangle \nonumber \\
&+& \bigg[ \frac{(1-\langle C_2 \rangle)^2}{(1-\langle C\rangle )^2}\bigg] \sqrt{\frac{j_2(2 j_2 - 1)}{(j_1 + j_2)(2(j_1 + j_2)-1)}} | j_1; j_2-2\rangle \nonumber\\
&+&\bigg[ \frac{2(1-\langle C_1 \rangle)(1-\langle C_2 \rangle)}{(1-\langle C\rangle )^2}\bigg] \sqrt{\frac{j_1 j_2}{(j_1 + j_2)(2(j_1 + j_2)-1)}} | j_1-1; j_2-1\rangle ~.
\end{eqnarray}
\vspace{-0.5cm}
\end{widetext}
Normalization of $|J_m, J_m - 2\rangle $ in Eq.\ref{Eq.75} provides Eq. \ref{Eq.76}
\begin{widetext}[!h]
\begin{eqnarray}\label{Eq.76}
 \bigg[ \frac{(1-\langle C_1 \rangle)^4}{(1-\langle C\rangle )^4}\bigg] \frac{j_1(2 j_1 -1)}{(j_1 + j_2)(2(j_1 + j_2)-1)} + \bigg[ \frac{(1-\langle C_2 \rangle)^4}{(1-\langle C\rangle )^4}\bigg] \frac{j_2(2 j_2 - 1)}{(j_1 + j_2)(2(j_1 + j_2)-1)} \nonumber \\+ \bigg[ \frac{4(1-\langle C_1 \rangle)^2(1-\langle C_2 \rangle)^2}{(1-\langle C\rangle )^2}\bigg] \frac{j_1 j_2}{(j_1 + j_2)(2(j_1 + j_2)-1)} = 1 ~.
\end{eqnarray}
\vspace{-0.5cm}
\end{widetext}
\end{enumerate}
In the same manner we can find the coefficients for all other states in this (highest) multiplet. 
The normalization conditions in the above states implies $\langle C_1\rangle = \langle C_2\rangle =\langle C\rangle$ . This condition is further consistent with normalization of all other states. This in turn implies that  the CG coefficients are independent of $\langle C_1 \rangle, \langle C_2 \rangle $ and $\langle C \rangle$.
CG coefficients for states in other multiplets then are obtained by using standard method and naturally are found to be independent of  $\langle C_1 \rangle, \langle C_2 \rangle $ and $\langle C \rangle$.\\
Now we proceed to find the coefficients  using $J_+$. Let us start with the initial lowest $m$ value state in the highest multiplet.  We obtain the following results\\
\begin{enumerate}
 \item  For ${ \bf  J = J_{m} , M= - J_{m} } $ ,  where $ J_m $ stands for $ J_{max} =  (j_1 + j_2)$:\\
\begin{eqnarray}\label{Eq.77}
|J_m , J_m \rangle = |m_1=-j_1~;~m_2=-j_2 \rangle ~.
\end{eqnarray}
The coefficient is $ \langle -j_1;-j_2~|~J_m, J_m \rangle = 1 $ by normalization.
\item For ${ \bf  J = J_{m} , M= -J_{m} + 1 } $ :\\
We get ,
\begin{eqnarray}\label{Eq.78}
|J_m, -J_m + 1\rangle =& \frac{(1-\langle C_1 \rangle)}{(1-\langle C\rangle)} \frac{\sqrt{j_1}}{\sqrt{j_1+j_2}} |-j_1+1;-j_2 \rangle \nonumber \\
+&\frac{(1-\langle C_2 \rangle)}{(1-\langle C\rangle)} \frac{\sqrt{j_2}}{\sqrt{j_1+j_2}} |-j_1;-j_2+1 \rangle ~.
\end{eqnarray}
On normalization of $|J_m, -J_m + 1\rangle $ in Eq.\ref{Eq.78} , we get the following condition:
\begin{eqnarray}\label{Eq.79}
\bigg[ \frac{(1-\langle C_1 \rangle)^2}{(1-\langle C\rangle )^2}\bigg] \bigg[\frac{j_1}{j_1+j_2}\bigg] + \bigg[ \frac{(1-\langle C_2 \rangle )^2}{(1-\langle C\rangle )^2} \bigg] \bigg[\frac{j_2}{j_1+j_2} \bigg] \nonumber\\
= 1 ~.
\end{eqnarray}
\item ${ \bf  J = J_{m} , M = -J_{m}+2 } $ 
which on calculation gives Eq.\ref{Eq.80}
\begin {widetext}[!h]
\begin{eqnarray}\label{Eq.80}
|J_m, 2-J_m \rangle &=& \bigg[ \frac{(1-\langle C_1 \rangle)^2}{(1-\langle C\rangle )^2}\bigg] \sqrt{\frac{j_1(2 j_1 -1)}{(j_1 + j_2)(2(j_1 + j_2)-1)}} | 2-j_1~;~-j_2\rangle \nonumber \\
&+& \bigg[ \frac{(1-\langle C_2 \rangle)^2}{(1-\langle C\rangle )^2}\bigg] \sqrt{\frac{j_2(2 j_2 - 1)}{(j_1 + j_2)(2(j_1 + j_2)-1)}} | -j_1~;~2-j_2 \rangle \nonumber \\
&+&\bigg[ \frac{2(1-\langle C_1 \rangle)(1-\langle C_2 \rangle)}{(1-\langle C\rangle )^2}\bigg] \sqrt{\frac{j_1 j_2}{(j_1 + j_2)(2(j_1 + j_2)-1)}} | 1-j_1~;~1-j_2\rangle ~.
\end{eqnarray}
\vspace{-0.5cm}
\end{widetext}
Again, on normalization, of $|J_m, -J_m + 2\rangle $ in Eq.\ref{Eq.80} , we get the condition given in Eq. \ref{Eq.81}.
\begin{widetext}[!h]
\begin{eqnarray}\label{Eq.81}
\bigg[ \frac{(1-\langle C_1 \rangle)^4}{(1-\langle C\rangle )^4}\bigg] \frac{j_1(2 j_1 -1)}{(j_1 + j_2)(2(j_1 + j_2)-1)} + \bigg[ \frac{(1-\langle C_2 \rangle)^4}{(1-\langle C\rangle )^4}\bigg] \frac{j_2(2 j_2 - 1)}{(j_1 + j_2)(2(j_1 + j_2)-1)} \nonumber \\
+ \bigg[ \frac{4(1-\langle C_1 \rangle)^2(1-\langle C_2 \rangle)^2}{(1-\langle C\rangle )^2}\bigg] \frac{j_1 j_2}{(j_1 + j_2)(2(j_1 + j_2)-1)} =1 ~.
\end{eqnarray}
\end{widetext}
\end{enumerate}
The solution of normalization conditions in Eqs \ref{Eq.79} and \ref{Eq.81} implies that
$\langle C_1\rangle = \langle C_2\rangle =\langle C\rangle$ . This condition is further consistent with normalization of all other states. This further implies that the  $\langle C_1 \rangle, \langle C_2 \rangle $ and $\langle C \rangle$ dependence is dropped out from the Eqs \ref{Eq.78} and \ref{Eq.80} and hence all  the CG coefficients are independent of $\langle C_1 \rangle, \langle C_2 \rangle $ and $\langle C\rangle$.

 The above results can be verified explicitly with specific examples. We have verified it for $ J_1 = \frac{1}{2} = J_2$, $ J_1 = 1 = J_2 $ and $J_1 = \frac{1}{2}, J_2 = \frac{3}{2} $. However, the results  for two spin 1/2 particles only are presented below. 
 \begin{eqnarray}\label{Eq.82}
 |1,1\rangle =& \bigg|\frac{1}{2};\frac{1}{2}\bigg\rangle\nonumber \\
 |1,0\rangle =& \frac{1}{\sqrt{2}} \left [ \frac{1-\langle C_1 \rangle}{1-\langle C \rangle}~\bigg|-\frac{1}{2};\frac{1}{2}\bigg\rangle +  \frac{1-\langle C_2 \rangle}{1-\langle C\rangle}~\bigg|\frac{1}{2};-\frac{1}{2}\bigg\rangle \right ]    \nonumber\\
 |1,-1\rangle =&\frac{(1-\langle C_1\rangle)(1-\langle C_2 \rangle)}{(1-\langle C\rangle)^2}~\bigg|-\frac{1}{2};-\frac{1}{2}\bigg\rangle \nonumber \\
 |0,0\rangle =& \frac{1}{\sqrt{2}} \left [ -\frac{1-\langle C_1\rangle}{1-\langle C\rangle}~\bigg|-\frac{1}{2};\frac{1}{2}\bigg\rangle +  \frac{1-\langle C_2 \rangle}{1-\langle C \rangle}~\bigg|\frac{1}{2};-\frac{1}{2}\bigg\rangle \right ] ~.
 \end{eqnarray}
 The normalization to the states $|1,0\rangle $ and $|1,-1\rangle$ provide condition similar to  Eq.\ref{Eq.74},\ref{Eq.76} and are now written as
\begin{eqnarray}\label{Eq.83}
(1-\langle C_1 \rangle)^2 + (1-\langle C_2 \rangle)^2 &=& 2 (1-\langle C \rangle)^2 \\
\label{Eq.84}
(1-\langle C_1 \rangle)^2 (1-\langle C_2 \rangle)^2 &=& (1-\langle C \rangle)^4 
\end{eqnarray}
Solving both the equations simultaneously, we get, $\langle C \rangle = \langle C_1 \rangle = \langle C_2 \rangle $
This condition is consistent with the normalization of the state $|0,0\rangle $. Also, if the starting state is considered as the normalized state $|J=1, M=-1\rangle = |-1/2 ; -1/2 \rangle$, we obtain from Eq.\ref{Eq.79}, \ref{Eq.81}, the same conditions as Eq.\ref{Eq.74}, \ref{Eq.76} and hence conclude the same condition on $C$, $C_1$,  $C_2$. With this information, we find the CG coefficients for the specific case where $j_1 =j_2 =1/2$ and $m_1$ (or $m_2$) $= \pm 1/2$  from Eq.\ref{Eq.73}, \ref{Eq.75}, \ref{Eq.78}, \ref{Eq.80} and list them in the table 1 \ref{Table1}.
\begin{widetext}[!h]
\vspace{-0.5cm}
\begin{table}[H]\label{Table1}
\begin{center}
\begin{tabular}{|p{3cm}||p{4cm}||p{4.5cm}||p{4.5cm}|}
 \hline
 \multicolumn{4}{|c|}{States} \\
 \hline
$|J , M \rangle $ & $ |m_1;m_2 \rangle $ & $ \mbox{Using } J_+ $ & $\mbox{Using } J_- $ \\[1.0ex]
 \hline
$|1,1 \rangle $  & $|1/2~; 1/2 \rangle $   &$ \frac{(1-\langle C_1 \rangle ) (1-\langle C_2 \rangle)}{(1-\langle C^2\rangle)} = 1$ & $ 1 $ \\[1.5ex]
 $ |1 ,0 \rangle $ & $|-1/2~; 1/2 \rangle $  & $ \frac{(1-\langle C_2 \rangle )}{\sqrt{2}(1-\langle C\rangle)}= \frac{1}{\sqrt{2}} $ & $ \frac{(1-\langle C_1 \rangle )}{\sqrt{2}(1-\langle C\rangle)} = \frac{1}{\sqrt{2}} $ \\[1.5ex]
  $|1 ,0 \rangle $ & $|1/2~; -1/2 \rangle $  & $ \frac{(1-\langle C_1 \rangle )}{\sqrt{2}(1-\langle C\rangle)}= \frac{1}{\sqrt{2}} $ & $ \frac{(1-\langle C_2 \rangle )}{\sqrt{2}(1-\langle C\rangle)}= \frac{1}{\sqrt{2}} $ \\ [1.5ex]
 $|1,-1 \rangle $  & $|-1/2~; -1/2 \rangle $   & $ 1 $ & $ \frac{(1-\langle C_1 \rangle ) (1-\langle C_2 \rangle)}{(1-\langle C^2\rangle)} = 1$ \\[1.5ex]
 \hline
\end{tabular}
\caption{Calculated CG coefficients for two spin 1/2 particles.}
\end{center}
\end{table}
\vspace{-1cm}
\end{widetext}

In case of standard QM, the CG-coefficients for a particular state are always same irrespective of the method of calculation, i.e, using $J_+$ or $J_- $. The same thing is true for GUP modified quantum mechanics also. This resolves the anomaly reported in \cite{PP2} where CG coefficients obtained by applying $J_-$ to the highest $J_m$ value state are different than that of when calculated by applying $J_+$ to the lowest $J_m$ value state in the same multiplet. We further showed that CG coefficients do not receive any correction due to quadratic order GUP and any angular momentum coupling remains unaffected even though the angular momentum algebra gets modified. Explicit examples were also considered to support our claim. Further this result can be generalized for  all composite systems in a straightforward manner. 
\section{Conclusion}
We have developed the operator formalism of SHO using the creation and annihilation operators in the framework of GUP. This formalism may find applications in wide variety of problems in QM one of which we have tried to address in the paper. Using this formalism and SMAM, we have derived GUP modified angular momentum algebra. While the GUP SHO may be useful in its own right, the angular momentum algebra developed may be used to investigate GUP  induced effects in Hydrogen atom, Quantum Rotor and other models involving the angular momenta operators. As such, a wide number of problems in atomic and molecular physics can be approached using the algebra developed. Further, using this GUP modified angular momentum algebra, we have studied the effect of GUP in the coupling of angular momentum for an arbitrary two particle composite system. We have also found the associated CG coefficients in the GUP modified systems. Some anomalies in calculating CG coefficient reported in the literature are removed in our formulation. Interestingly, we observe that CG coefficients receive no corrections up to quadratic GUP. This has been verified explicitly for two particle systems. The coupling of angular momentum remains unaffected to quantum gravity effects is indeed an important result and may have very important consequences in the study of Black holes and many other astro-physical problems where gravity plays an important role.

\end{document}